\documentstyle[11pt,ysc,twoside,epsf]{article}
\def\edcomment#1{\iffalse\marginpar{\raggedright\sl#1\/}\else\relax\fi}
\reversemarginpar

\begin{document}
\title{The Light Curve Variations of The Active Binaries With Hot Subdwarf Component}

\author{Esin S{\.I}PAH{\.I}}
\affil{Ege University Observatory, 35100, Bornova, {\.I}zmir, TURKEY }

\author{Serdar EVREN}
\affil{Ege University Observatory, 35100, Bornova, {\.I}zmir, TURKEY}

\begin{abstract}
We present the light curve variations of the two active binaries with hot subdwarf component.
According to the brightness variations outside of the eclipses, the giant 
components of the systems are chromospherically active stars. The dark and cool
active structures on this components cause the variations of the total light of the 
systems.
\end{abstract}

\section{Introduction}

Nowadays, there are few systems containing a hot subdwarf and a cool
giant component. The observations of these systems are very
important for the evolution of the hot subdwarf, studying the
activity of cool giant stars and also for testing the evolution
models. Hot subdwarf stars are the helium burning stars (the mass of
He core is about 0.5 solar masses) covered with a very thin hydrogen
shell (about 0.02 solar masses). They form a relatively narrow
sequence at the blue end of the horizontal branch (HB). Therefore
they are referred to as Extended Horizontal Branch Stars (EHB). The
major difference to normal HB stars is the fact that they do not
evolve to the asymptotic giant branch. Many of them are members of
binary systems with late-type components. The secondary components
which we observed are cool (G8 or K0 III-IV) and shows solar-like
activity. We give some results of the photometric studies of V1379
Aql and FF Aqr. The observations of the systems were carried out
with the 48 cm Cassegrain telescope at Ege University Observatory in
2003 year.

\section{The Light Curve Variations}

V1379 Aql is a binary system containing a red giant (K0 III/IV) and
a hot subdwarf B star. The first indication of chromospheric
activity on the giant star came from the detection of Ca II H \& K
emission by Bidelman {${\&}$} MacConnell (1973). Photometric
variations with an amplitude of about 0$^{m}$.2 were first observed
by Henry et al. (1982). The system is asynchronous, the rotational
period of 25.4 days found by Balona et al. (1987) being longer than
the orbital period (20.7 days) determined from radial velocity
measurements (Balona 1987, Fekel et al. 1993). Hooten \& Hall (1990)
determined a photometric period of about 26 days with a variation
amplitude of 0$^{m}$.20-0$^{m}$.25 in the V band. In this study, the
photometric period of the system was found 25.7 days with an
amplitude of 0$^{m}$.13 in the V band. This might suggest that the
photometric period of the system is changing. Also, the amplitude of
the light in V band is different from another data which are given
previous years.

FF Aqr is an eclipsing binary containing a hot subdwarf OB star and
a chromospherically active companion as G8 III star. The light curve
which obtained by Dworetsky et al. (1977) in 1975 displays an
asymmetrical wave distortion outside of the eclipse with the maximum
at about 0$^{P}$.55 and an amplitude of 0$^{m}$.35 in V band. Dorren
et al. (1983) observed the system in the b, v, y Str\"{o}mgren
bands. The observations show that the wave maximum had moved to
phase 0$^{P}$.72. Marilli et al. (1995) presented the photometry of
the system. In this study, the maximum of the wave distortion
outside of the eclipse was found to occur at about 0$^{P}$.5 with an
amplitude of 0$^{m}$.12. The light curve of the system in V band was
found to be changed when it was compared with another light curves
given by Dworetsky et al. (1977), Dorren et al. (1983) and Marilli
et al. (1995). The maximum of the wave outside of the eclipse
occurred at different phases. This is indicative of the spotted star
surface.

\begin{figure}
\plotone{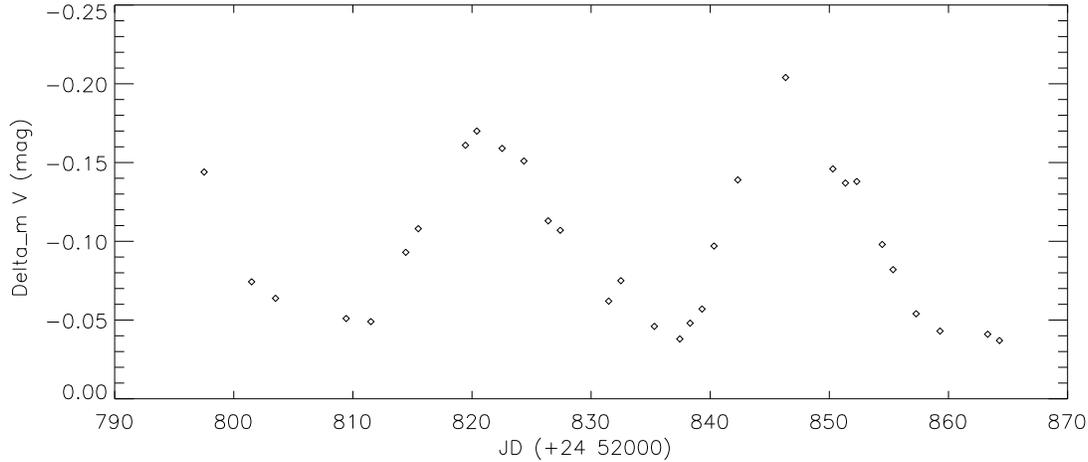}
\caption{The V light curve against Julian Date of V1379 Aql.}
\end{figure}

\begin{figure}
\plotone{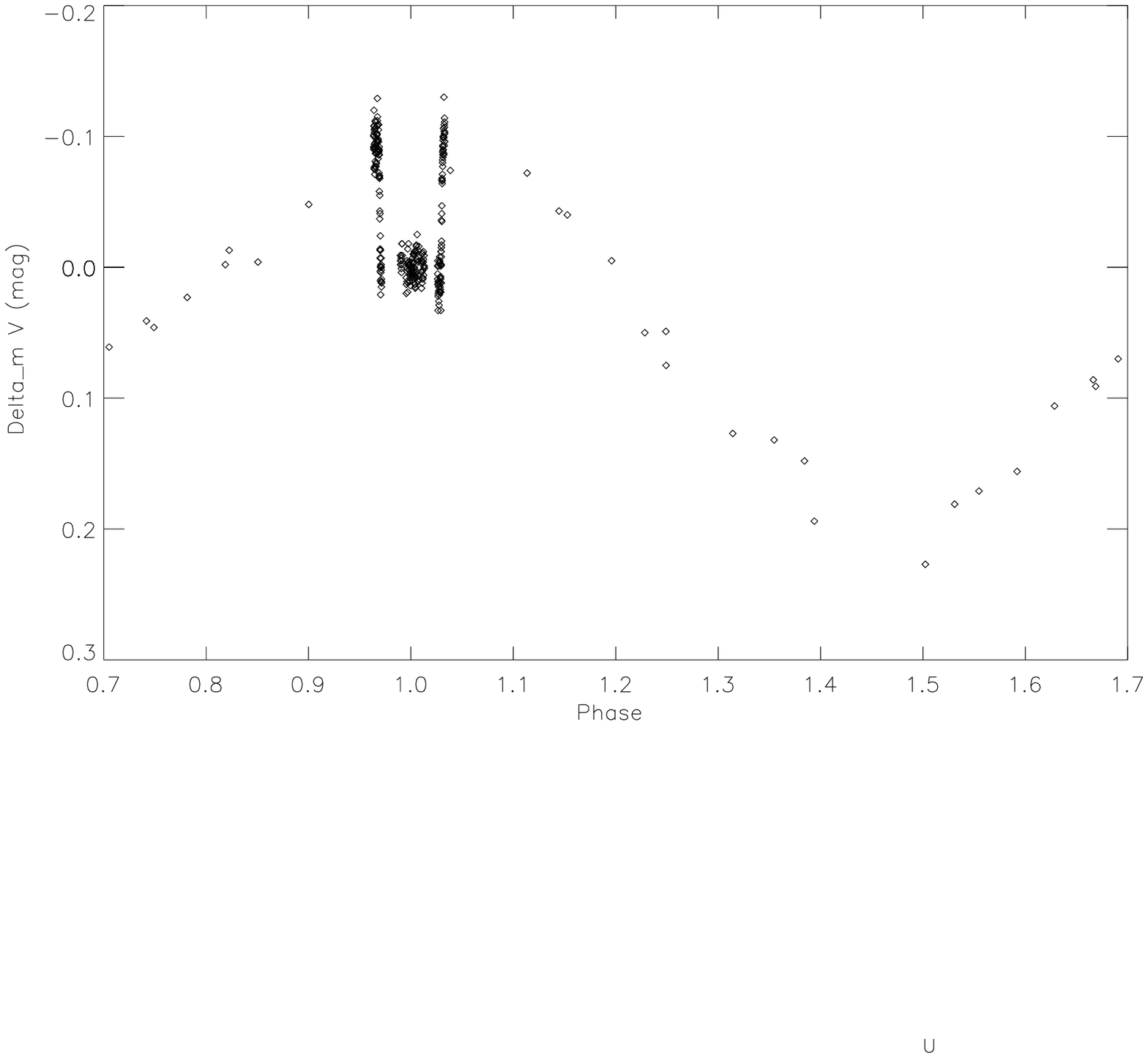}
\caption{The V light curve of FF Aqr from 2003 observations.}
\end{figure}

\section{Conclusion}

The shape of the chromospherically active stars' light curves
depends on the parameters of active regions such as cool spot or
spot groups. It varies with time depending on the location (latitude
and longitude), numbers, relative sizes and relative temperatures of
the active regions with respect to the unspotted photosphere,
life-times and the orientation of the rotation axis. Any change in
some of these variables in short or long time interval will affect
the shape and the amplitude of the light curve with time. Long-term
continuous monitoring of a spotted star covering at least ten years
may reveal better understanding of the behavior of surface
inhomogeneities, i.e. activity cycles, migration of the active
regions, changes of the photometric periods and surface differential
rotation. The Sun's differential rotation has been known for a long
time by interpreting location of the sunspots. At present we know
from the butterfly diagram that the active regions to formation at
the middle latitudes at the beginning of the 11-year activity cycle
and the activity get stronger with increasing spot numbers. When the
active regions appear close to the equator the activity decreases
and the cycle is ending. The present technique is inadequate to
resolve stellar disks. Therefore we could reveal surface
differential rotation from the migration of starspots. In other
words, presence of surface differential rotation in the stars can be
inferred from changes in rotational period (Hall 1991).

\section{Acknowledgements}
This work has been partly supported by the Research Foundation of Ege University with the project number 2003/FEN/003.

\begin {references}
\reference Bidelman W.P., MacConnell D.J. 1973, AJ, 78, 687
\reference Henry G.W., Murray S., Hall D.S. 1982, IBVS 2215
\reference Balona L., Lloyd Evans T., Simon T., Sonneborn G. 1987, IBVS 3601
\reference Hooten J.T., Hall D.S. 1990, ApJS, 74, 225
\reference Hall D.S. 1991, ApJ, 380L, 85
\reference Fekel F.C., Henry G.W., Busby M.R., Eitter J.J. 1993, AJ, 106, 2370
\reference Marilli E., Frasca A., Bellina Terra M., Catalano S. 1995, A\&A, 295, 393
\end {references}
\end{document}